# Detection of Malicious Websites Using Machine Learning Techniques

Adebayo Oshingbesan
*Carnegie Mellon University*
Kigali, Rwanda
oaadebay@andrew.cmu.edu

Aime Munezero
*Carnegie Mellon University*
Kigali, Rwanda
amunezer@andrew.cmu.edu

Chukwemeka Okobi
*Carnegie Mellon University*
Kigali, Rwanda
cuo@andrew.cmu.edu

Kagame Richard
*Carnegie Mellon University*
Kigali, Rwanda
krichard@andrew.cmu.edu

Courage O Ekoh
*Carnegie Mellon University*
Kigali, Rwanda
coekoh@andrew.cmu.edu

*Abstract*—In detecting malicious websites, a common approach is the use of blacklists which are not exhaustive in themselves and are unable to generalize to new malicious sites. Detecting newly encountered malicious websites automatically will help reduce the vulnerability to this form of attack. In this study, we explored the use of ten machine learning models to classify malicious websites based on lexical features and understand how they generalize across datasets. Specifically, we trained, validated, and tested these models on different sets of datasets and then carried out a cross-datasets analysis. From our analysis, we found that K-Nearest Neighbor is the only model that performs consistently high across datasets. Other models such as Random Forest, Decision Trees, Logistic Regression, and Support Vector Machines also consistently outperform a baseline model of predicting every link as malicious across all metrics and datasets. Also, we found no evidence that any subset of lexical features generalizes across models or datasets. This research should be relevant to cybersecurity professionals and academic researchers as it could form the basis for real-life detection systems or further research work.

I. INTRODUCTION

The internet has seen a spike in usage in recent years, as several entities use it for a variety of reasons ranging from personal to business use. Unfortunately, some entities are capitalizing on the ubiquitousness of the internet to carry out malicious activities. A common mechanism through which such malicious activities can be carried out is via a malicious website. A malicious website is a site that attempts to install malware onto users' devices to carry out some unauthorized activities such as stealing financial information, having access to confidential data such as passwords among others [1 - 3]. Thus, discovering these malicious URLs has been of great interest in the field of cybersecurity and is our interest in this research work.

One common technique in the identification of malicious URLs is the use of a blacklist. While blacklisting a URL has been effective to some extent, the fact that these URLs are rapidly evolving means blacklists are not sufficient defense against this form of attack. Machine learning techniques have been proposed as an effective tool in tracking malicious URLs as they could be used to find malicious websites even if they have never been seen before unlike a blacklist [4]. Contextually, machine learning (ML) is the study of computer algorithms that improve automatically through experience and using data [5]. Since ML models can understand the underlying lexical structure of URLs, they give better insights into classifying URLs [6 -7].

In this research, we aim to answer the question - are there machine learning techniques that have consistently high performance (i.e., have an average position of 3 or less) across several datasets using lexical features? Our initial hypothesis is that tree-based models will perform consistently better than other types of models. This research should be relevant to cybersecurity professionals and academic researchers as it could form the basis for real-life detection systems or further research work. Our research objective is to understand the generalization capabilities of various machine learning models for the classification of malicious URLs. Specifically, this research scope only covers using lexical features and common machine learning algorithms. More advanced features such as DNS-based features, content-based features, and so on are beyond the scope of this research. Similarly, advanced machine learning techniques such as deep learning and reinforcement learning will not be studied. Our research assumptions are:

*1) There exist open-access datasets with human-labeled classifications of the state of maliciousness or benignity of URLs contained in these datasets.*
*2) These open-access datasets are representative samples of the kind of URLs that are encountered in the wild.*
*3) Lexical features are representative of the state of maliciousness or benignity of a website. At the very least, they are better than random guessing.*
*4) Machine learning modeling is a sufficient modeling technique for this problem to learn the feature space and make correct predictions.*

II. IMPORTANCE AND PRIOR WORK

Various researchers have proposed the use of machine learning, data mining, and even deep learning techniques to detect malicious websites. The success of these techniques highly depends on the quality and combination of relevant characteristic features of web pages [8] which may include network traffic information, content characteristics, lexical features of URLs, and even domain name system (DNS) information. However, extracting these attributes may be costly and sometimes require downloading complete web pages [9] or looking up various DNS servers and ISPs to get enrichment data like geo-location, registration records, and network information [10], which face the problem of network latency, making them impractical for real-time systems [7, 11]. Due to the difficulties of using non-lexical features to detect malicious URLs in real-time despite their high accuracy values [12-13], previous works have explored the use of URL lexical features only and it has been proven that URL features alone can produce an accurate means of detecting malicious webpage in real-time systems [7, 9, 13].

Even in systems that use lexical-based features only, the performance reported across the literature is also highly dependent on the model and dataset used [7, 9, 13 - 17]. For example, in [14], the authors compared three supervised machine learning models (k-nearest neighbor, support vector machine (SVM), and naive bayes classifier) and two unsupervised machine learning models (k-means and affinity propagation). The result of the analysis found that supervised models outperformed unsupervised machine learning models by a small margin. Furthermore, the size of the dataset may also pose a problem. For example, in [16], association rule mining was used in combination with several machine learning models to classify URLs as either malicious or benign based on URL-derived features. To deal with the class imbalance and small dataset size, the authors employed the Synthetic Minority Over-Sampling Technique (SMOTE). The evaluation of the models before and after class balancing reveals that after class balancing and the subsequent increase in dataset size, most models experienced significant improvements.

While several studies have utilized popular machine learning techniques to classify URLs as malicious with some of these studies doing a comparative analysis of ML models on their dataset, a major flaw still exists since these analyses are usually dataset-specific. Thus, the critical goal of this research work is understanding how machine learning models generalize over several datasets to avoid spurious results that arise from the dataset used rather than the analysis performed as seen in some previous research works in other fields [18-20]. Our major contribution will be helping security experts and academic researchers understand what works best in a wide range of scenarios without compromising on essential qualities of malicious website detection systems such as speed, high accuracy, low false-negative rates, and so on.

III. METHODOLOGY

The research problem can be subdivided into five main sub-problems: data collection, data pre-processing, lexical feature engineering, machine learning modeling, cross-datasets analysis.

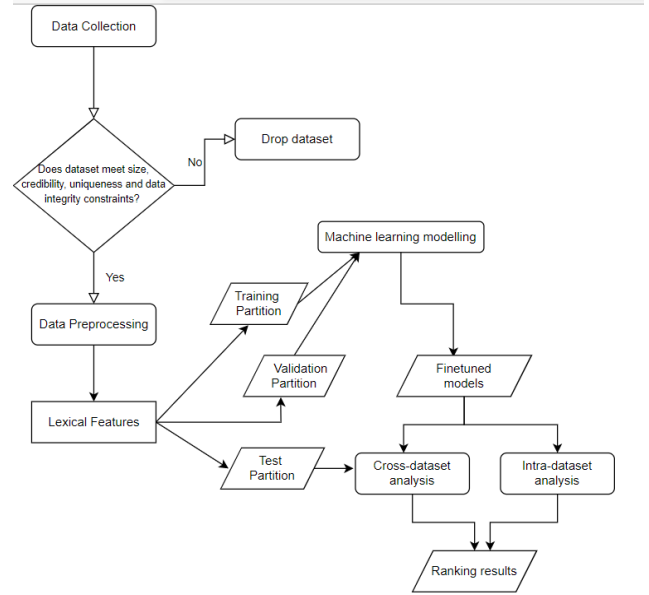

Fig. 1 The interconnection between sub-problems

A. Data Collection

This involved gathering the datasets that were used in this research. The sub-steps were:

*1)* Manually searching through online datasets repositories such as datasetsearch.google.com and Kaggle.com to find potential datasets that could be used for analysis.

*2)* Pruning datasets using metrics such as size, uniqueness, and the credibility of the source. Uniqueness is measured by a maximum of 20% overlap with the other datasets used in this analysis. The credibility of a source is measured by verifiable explanations of how the dataset was curated from. This would give a shortlist of datasets that would go through data integrity checking.

*3)* Performing data integrity checking by manually verifying the labels of a random subset of size 100 for each of the datasets. Any dataset that failed this manual verification of 80% accuracy as ascertained by a human verifier was dropped.

B. Data Pre-processing

This entailed standardizing the datasets to ensure they all have the form – URL (a string) and label (benign or malicious). This stage involved writing python scripts that do this standardization based on the dataset that we are dealing with. Using these scripts, each of the datasets will then be standardized for analysis.

C. Lexical Feature Engineering

This involved generating values for the feature space of this analysis using the URL lexical properties. Lexical feature engineering had the following steps:

*1)* Searching literature to obtain the lexical features used in this kind of analysis.

*2)* Ranking these features in terms of popularity, literature importance, and novelty.

*3)* Designing at least one entirely new feature based on our domain knowledge.

*4)* Writing python scripts that take in the standardized dataset and return all the engineered features that will then be used for further analysis.

*5)* Partition of the datasets into training partition (34%), validation partition (33%), and test partition (33%).

*D. Machine Learning Modelling*

This was the heart of our analysis. It involved converting the standardized dataset into relevant models. This sub-problem involved:

*1)* Shortlisting machine learning models based on literature use for malicious website detection or similar tasks.

*2)* Pruning the shortlisted machine models based on empirical performance both in the literature related to malicious website detection and other well-known tasks to 10 models.

*3)* Determining what metrics would be used for training and validating the models.

*4)* Building the machine learning models, training them on the training partition, and then carrying out validation with hyperparameter optimization on the validation set, to ensure that the models are well fine-tuned to the task at hand.

*5)* Saving these fine-tuned models for cross-dataset analysis.

*E. Cross-Datasets Analysis*

The cross-datasets analysis involved testing the models on the test partition and performing comparative analysis on all the models' performances across every dataset in the test partition. To guarantee that our modeling choices were appropriate, we used the same models to train, validate and test on one of the datasets in a single dataset analysis. If the results of the single dataset analysis were okay, it meant that our modeling choices were solid, and the results obtained from the cross-datasets analysis were justifiable. Ranking tables were obtained to show the models that generalized best across the datasets based on the comparative analysis using the mean rank score.

## IV. RESULTS AND DISCUSSION

*A. Data Collection*

At the end of the data collection process, we ended up with 16 datasets with sizes ranging from 20 thousand URLs to 600 thousand URLs, totaling over 2 million URLs. We dropped duplicate URLs to ensure that every URL is unique. The duplicate removal process ensures that the only URL overlap present in the dataset were URLs that had different lexical forms but may refer to the same site such as www.google.com, https://www.google.com, and google.com. We left these URLs in because of their lexical variety which would guarantee robustness.

*B. Data Pre-processing*

Since the datasets were from different sources, their representations differ. One dataset may represent malicious URLs as 1 in the label category while another may put a string of malicious or bad as the label. Thus, we standardized all the 16 datasets such that we had a URL column containing the URL string and a label containing 0 for benign URLs and 1 for malicious URLs. In total, 30% of the datasets were malicious with slight variations across the datasets.

*C. Lexical Feature Engineering*

Using literature data, we were able to come up with 78 lexical features ranging from features that describe counts such as counts of special characters (such as &, #, @, $ and so on) to count of alphabets/numbers to the data on the length of the hostname, the top-level domain, how many paths were in the URL among others. We also had 300 word2vec-based features based on the novel contribution of [7]. Furthermore, we built on the work of [7] to propose two novel features – benign score and malicious score. Rather than use an n-gram model built on blacklist words as they did, we built two "language models" for malicious and benign URLs. Language modeling (LM) is the use of various statistical and probabilistic techniques to determine the probability of a given sequence of words occurring in a sentence in that language [21]. In this analysis, we tried to determine the probability of a sequence of characters occurring in a URL string given that the URL is benign or malicious. The intuition is that if a URL is malicious, the malicious URL language model would give the URL string a higher probability of occurrence while the benign URL language model would give the URL a lower probability of occurrence. The converse would be true if the URL is benign. Together, these two scores should be predictive of the malicious state of a given URL. After all the feature engineering, we ended up with 380 features that we believed should be highly predictive of whether a URL is malicious or not.

*D. Machine Learning Modelling*

Our final models consist of six groups of classifiers - eight supervised models and two unsupervised models:
- linear models (Logistic Regression (LR), Linear Support Vector Machines (SVM)),
- tree-based models (Decision Tree (DT), Random Forest (RF), Categorical Boosting (CB)),
- neighbors-based model (K-Nearest Neighbours (KNN)),
- neural model (Feed Forward Neural Network (FFNN)),
- statistical models (Naïve Bayes (NB))
- unsupervised models (KMeans and Gaussian Mixture Model (GMM))

We also chose five metrics for comparative analysis - Area under the Receiver Operating Curve (AUC-ROC), Recall (REC), Precision (PCSN), F1, and Accuracy (ACC). The train, validation, and test data contained 6, 5, and 5 unique datasets, respectively. The training and validation process involved other processes like feature selection, feature scaling, dimensionality reduction, and hyperparameter tuning. Feature selection showed that all the ten models found that the word2vec based features were not predictive of maliciousness. We believe this is because word2vec models are usually dataset-specific. Thus, they are not as useful in a cross-dataset analysis. However, eight of ten models found our novel features to be important in the prediction, showcasing their clear value. Lastly, no group of features dominated the prediction choice of any of the models across all datasets.

*E. Cross-Datasets and Single Dataset Analysis*

Table 1 shows the results obtained from the single dataset analysis. Table 2 shows the average results obtained from the cross-dataset analysis. Table 3 shows the models' average rank performance across the five test datasets (DS1 – 5) using the average ranks obtained for each metric per dataset for the combined dataset analysis. We define rank performance (RNK) to be equal to the model's ranked position in comparison to other ML models if the model performs better than the baseline model across all the metrics that or 10 otherwise. We chose our baseline model to be a dumb classifier that assumes every link it comes across as malicious. The average rank performance is then the mean rank performance rounded to the nearest integer.

TABLE I. TABLE OF METRIC SCORES FOR SINGLE DATA SET ANALYSIS

| F1 | AUC-ROC | ACC | REC | PCSN |
|---|---|---|---|---|
| 0.78 | 0.83 | 0.85 | 0.73 | 0.84 |

TABLE II. TABLE OF AVERAGE METRIC SCORES ACROSS ALL DATASETS FOR CROSS_DATASET ANALYSIS

| DATASET | METRICS | | | | |
| | F1 | AUC-ROC | ACC | REC | PCSN |
|---|---|---|---|---|---|
| DS 1 | 0.39 | 0.58 | 0.52 | 0.68 | 0.38 |
| DS 2 | 0.39 | 0.55 | 0.55 | 0.43 | 0.71 |
| DS 3 | 0.39 | 0.55 | 0.55 | 0.43 | 0.71 |
| DS 4 | 0.39 | 0.55 | 0.55 | 0.44 | 0.71 |
| DS 5 | 0.39 | 0.55 | 0.55 | 0.44 | 0.70 |

TABLE III. TABLE OF AVERAGE MODEL RANK PERFORMANCE ACROSS DATASETS

| MODELS | DATASETS | | | | | |
| | DS 1 | DS 2 | DS 3 | DS 4 | DS 5 | RNK |
|---|---|---|---|---|---|---|
| KNN | 2 | 1 | 2 | 1 | 1 | 1 |
| SVM | 3 | 5 | 4 | 4 | 4 | 4 |
| RF | 10 | 4 | 4 | 4 | 4 | 5 |
| DT | 5 | 7 | 6 | 6 | 6 | 6 |
| LR | 6 | 6 | 7 | 6 | 6 | 6 |
| NB/CB/ GMM/ KMEANS/ FFNN | 10 | 10 | 10 | 10 | 10 | 10 |

From the table I, we can see from the performance of the models on the single dataset analysis was good even though we finetuned the models for the cross-dataset analysis only. This means that our modeling choices are valid and cannot be the reason for any bad performance obtained on the cross-dataset analysis. Table II shows that the models' average performance deteriorates rapidly when cross dataset analysis is performed, implying that they struggle to replicate their performance in a cross-dataset analysis.

Contrary to our initial hypothesis, table 3 shows that only KNN had an average rank equal to or lower than 3. However, we should still note that two out of the three tree-based models chosen for this analysis still finished in the top 5 showing their capabilities to generalize as originally hypothesized. The rankings in table III also show that unsupervised learners, simple models such as NB, and complex models such as CB and FFNN failed to generalize across datasets as they only had performance comparable to our baseline models across all the test datasets.

V. CONCLUSION

Based on our analysis, we conclude that the model that generalizes best across multiple datasets using lexical features is KNN. SVM, Random Forest, Logistic Regression, and Decision Trees are expected to perform better than the baseline model of assuming every link is malicious. Thus, the popular choice of tree-based models and linear models in literature for malicious links detection in single dataset analysis seems justified. However, as shown in our analysis, their performance in a cross-dataset analysis is not as good as that in a single dataset analysis. Thus, care must be taken in accepting their performance when reported as there are no guarantees that this performance will remain the same when tested on links that did not belong to the original training/validation set.

KNN seems to be the only model that performs adequately across datasets based on our analysis. This may be because of its nearest neighbor paradigm. Further concrete analysis needs to be carried out to investigate why. Also, we found no evidence that a specific subset of lexical features generalizes over datasets. Still, this could have been influenced by our processing choices. Hence, future works could investigate if this holds using different processing choices. Finally, since we have used only lexical features in this work, there is no evidence to suggest this analysis will generalize to other commonly used features such as content-based or DNS-based features. Consequently, future works could extend this analysis to other feature groups, either individually or in combinations with one another.